# In-Situ Single Particle Reconstruction Reveals 3D Evolution of PtNi Nanocatalysts During Heating

*Yi-Chi Wang[1,2,3], Thomas J. A. Slater[4]\*, Gerard M. Leteba[5], Candace I. Lang[5], Zhong Lin Wang[1,6], Sarah J. Haigh[2]\**

1. Beijing Institute of Nanoenergy and Nanosystems, Chinese Academy of Sciences, Beijing, 101400, China

2. Department of Materials, University of Manchester, Manchester M13 9PL, UK

3. School of Materials Science and Engineering, Tsinghua University, Beijing 100084, China

4. Cardiff Catalysis Institute, School of Chemistry, Cardiff University, Cardiff CF10 3AT, UK

5. Centre for Materials Engineering, Department of Mechanical Engineering, University of Cape Town, Cape Town 7700, South Africa

6. School of Materials Science and Engineering, Georgia Institute of Technology, Atlanta, GA 30332-0245, USA

Thomas J. A. Slater, Email: slatert2@cardiff.ac.uk

Sarah J. Haigh, Email: sarah.haigh@manchester.ac.uk

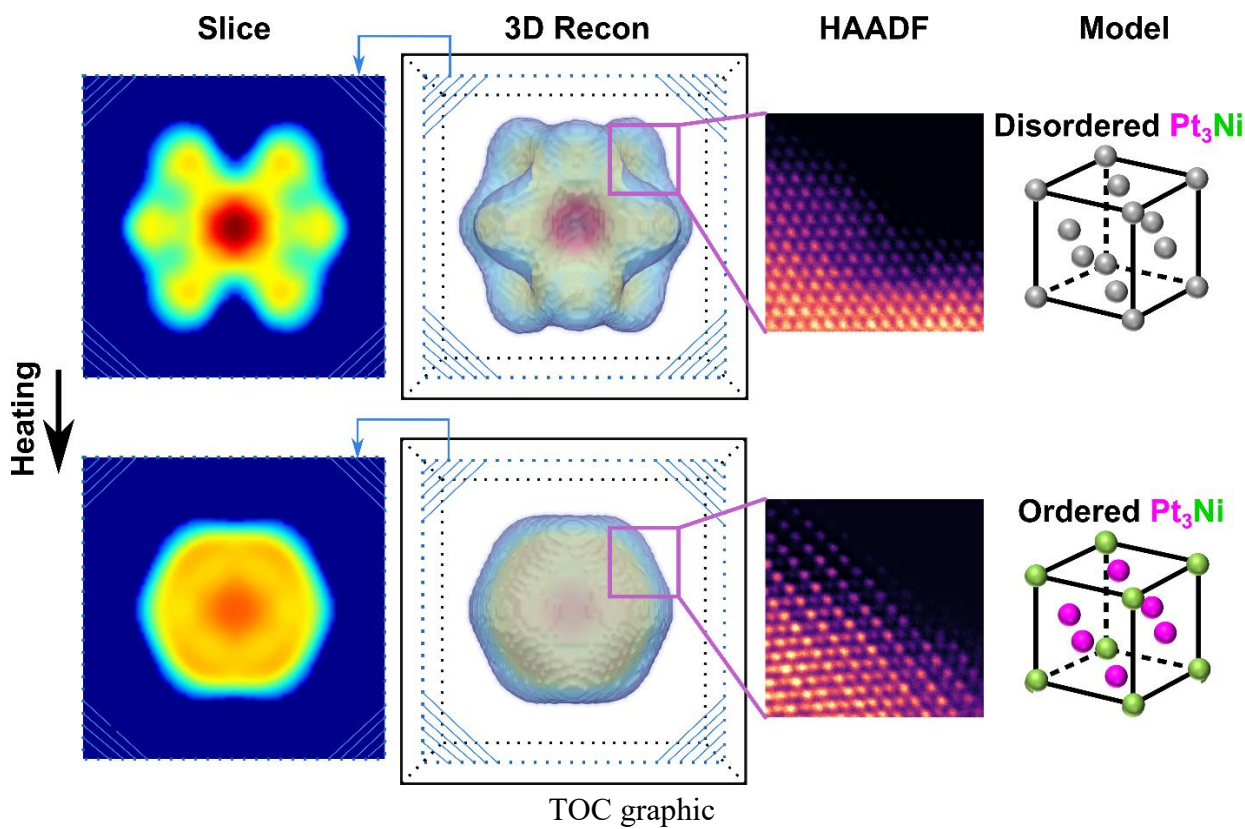

TOC graphic

ABSTRACT

Tailoring nanoparticles' composition and morphology is of particular interest for improving their performance for catalysis. A challenge of this approach is that the nanoparticles' optimized initial structure often changes during use. Visualizing the three dimensional (3D) structural transformation in-situ is therefore critical, but often prohibitively difficult experimentally. Although electron tomography provides opportunities for 3D imaging, restrictions in the tilt range of in-situ holders together with electron dose considerations limit the possibilities for in-situ electron tomography studies. Here, we present an in-situ 3D imaging methodology using single particle reconstruction (SPR) that allows 3D reconstruction of nanoparticles with controlled electron dose and without tilting the microscope stage. This in-situ SPR methodology was employed to investigate the restructuring and elemental redistribution within a population of PtNi nanoparticles at elevated temperatures. We further examined the atomic structure of PtNi and found a heat-induced transition from a disordered to an ordered phase. Changes in structure and elemental distribution were linked to a loss of catalytic activity in the oxygen reduction reaction. The in-situ SPR methodology employed here could be extended to a wide range of in-situ studies employing not only heating, but gaseous, aqueous or electrochemical environments to reveal in-operando nanoparticle evolution in 3D.

Platinum based nanoparticles are essential electrocatalysts in proton exchange membrane fuel cells (PEMFCs)[1] and when alloyed with 3d transition metals, exhibit enhanced functionalities for the oxygen reduction reaction (ORR).[2,3] The nanoparticles' morphology can also be tailored to enhance electrocatalytic performance, with high surface area and specific surface facets used to generate a larger number of specific active sites.[4–7] The result is alloyed Pt nanoparticles with controlled morphology that are exceptionally active for the ORR.[4–7]

However, synthesizing such complex nanoparticles is challenging; often requiring post-synthesis heating steps to tune the PtNi nanoparticles' morphology and surface composition.[8] Additionally, the working conditions experienced in PEMFCs can cause significant changes to the shape of high-surface area nanoparticles during use, with a negative effect on their long-term performance.[9] For example, the well-defined shape of octahedral PtNi nanoparticles has been seen to degrade after electrochemical cycling.[10] In-situ characterization methods are required to better understand nanoparticle evolution and thereby optimize the materials' post synthesis treatment and minimize in-service structural degradation.

The transmission electron microscope (TEM) is a powerful instrument to investigate nanoparticle structure, composition, and oxidation state down to the atomic scale. Although conventional TEM imaging, or scanning transmission electron microscope (STEM) imaging, is performed in static, high vacuum and room temperature environments, in-situ holders (which can apply heat, gaseous, or liquid environments) have now become a mature technology.[11,12] These holders complement the environmental TEM (ETEM) approach where a differentially-pumped objective lens allows low pressure gas to surround the sample.[13,14] These platforms provide opportunities to perform in-situ S/TEM characterization of nanoparticle size, shape and composition changes under operando conditions.

However, the vast majority of in-situ S/TEM studies of metal nanoparticles use only two-dimensional (2D) projected images, which cannot unambiguously resolve complex structural and compositional changes due to overlapping information along the third spatial dimension (parallel to the electron beam viewing direction). To remove the ambiguity of 2D projections, three-dimensional (3D) S/TEM imaging can be performed using electron tomography. Conventional tilt-series electron tomography requires imaging the same specimen many times, tilting the holder over a range of angles (often in the range ±75°) and collecting data at each tilt angle. The literature contains a number of reports of electron tomography studies of nanoparticles compared before and after ex-situ heating.[15,16] For example, Zhou et al. performed atomic-resolution STEM high-angle annular dark field (HAADF) tomography and obtained a 3D reconstruction of the same FePt nanoparticle after ex-situ heating in an oven at 520°C for 9, 16 and 26 minutes respectively.[15] They tracked the evolution of the internal atomic structure at these three heating times and found nuclei for the solid-solid phase transitions undergoing growth, shrinkage or dimensional fluctuation.

The disadvantages of such ex-situ heating studies include loss of information for intermediate steps, the challenge of locating identical nanoparticles before and after heating, and the potential for artefacts introduced during transfer from the furnace to the TEM. However, ex-situ studies are often necessary as commercial in-situ holders are generally not suitable for tomographic imaging studies due to the limitations on tilt range imposed by the small pole piece gap of the electron microscope. Specialized in-situ heating holders have been produced with narrower cross sections so that they can achieve higher tilt ranges, facilitating a number of in-situ tomographic heating studies.[17–19] For example, Epicier and co-workers have performed electron tomography using fast bright field (BF) imaging and a commercial heating holder in an ETEM to study the thermally induced structural changes of a series of supported nanocatalysts.[20,21] In-situ HAADF tomography has also been used to study the overall structural change[17,18] or alloying[19] of metal nanoparticles on heating in vacuum.

Environmental in-situ gas and liquid cells holders are bulkier than heating holders, meaning that reducing their lateral dimensions is more challenging (with most limited to a tilt range of ±25°), which generally prohibits conventional tilt-series tomography. A notable exception is the custom designed microchip liquid-cell used by Dearnaley *et al.* to achieve 3D information on the interaction between a bacteriophage and its host bacterium, albeit at a relatively poor spatial resolution.[22] An alternative tomographic approach removes the need to rotate the in-situ cell by exploiting the nanoparticles' tendency to undergo uncontrolled rotation in the liquid cell environment, Park and co-workers have shown that when coupled with the atomic resolution achievable by using a graphene liquid cell, the technique allows the 3D atomic structure of Pt nanoparticles to be reconstructed.[23,24] Nonetheless, for many nanoparticle systems and realistic environmental imaging conditions the factor that ultimately prevents 3D information being resolved is the sensitivity of the sample to the electron beam.

In structural biology, 3D imaging of proteins and macromolecules is achieved by averaging low electron dose images from many similar objects using the single particle reconstruction (SPR) technique.[25] This has also been demonstrated for inorganic nanoparticles[26], where the potential to fractionate electron dose has been shown to open up new possibilities for spectroscopic 3D imaging.[27] For in-situ imaging applications, the SPR approach has the advantage that it removes the need for a specialized high-tilt holder and so is compatible with commercial in-situ gas flow, liquid flow and electrochemistry holder platforms. Furthermore, the time-consuming nature of tilt-series based in-situ 3D imaging methods means that they can only track the evolution of a limited number of particles (often only one). Consequently, there can be questions regarding whether the transformation identified is applicable to a larger nanoparticle population. The SPR approach has the advantage that a far larger number of nanoparticles are included in the analysis.

Here, we demonstrate an in-situ 3D imaging SPR methodology and its application to reveal the 3D structural evolution of PtNi nanoparticles during thermal treatment. Using atomic resolution STEM-HAADF imaging, we additionally identify a phase transition from a chemically disordered phase to an ordered phase via heat treatment at 400°C. We compare the nanoparticles' catalytic performance after heat treatment and provide evidence for the importance of the support material in determining the nanoparticles structural evolution at a fixed temperature.

### Analysis of individual nanoparticle 3D morphology based on 2D projections

To exemplify the potential of the novel in-situ SPR approach, we investigated the thermal evolution of PtNi nanoparticles synthesized with a rhombic dodecahedral morphology, which possess a high activity for the ORR (see experimental methods for details of nanoparticle synthesis).[5] Figure 1a shows an example of the STEM-HAADF images used as inputs for the in-situ SPR analysis. Multiple regions were imaged at each temperature step to observe a sufficient number of randomly oriented PtNi nanoparticles. In total, 60 nanoparticles were characterized by STEM-HAADF (Figure 1) and STEM-EDX spectrum imaging (SI Section 1.3). A sparse distribution of nanoparticles on the SiN grid was employed (Figure 1a) to ensure an individual nanoparticle's evolution behavior was not affected by its neighbors (see SI Section 1 and 2 for further acquisition details). As preliminary tests showed these nanoparticles do not change significantly between room temperature and 200°C (SI Figure S1), this was used as the lowest temperature step, increasing from 200°C to 550°C with intervals of 50°C. Each temperature was held for 10 minutes to allow nanoparticles to restructure and then the sample was quenched to room temperature for image acquisition. The projected area of the nanoparticles at 200°C shows a narrow size distribution (SI Figure S1). The particles have previously been shown to have a homogenous population in terms of size and morphology suitable for room temperature SPR.[27] Each individual nanoparticle was identified (Figure 1b) and segmented (Figure 1c) to remove background intensity (further details of image processing steps can be found in SI Section 1).[28,29]

Figure 1d shows three examples of the structural evolution for extracted individual nanoparticles as a function of temperature.

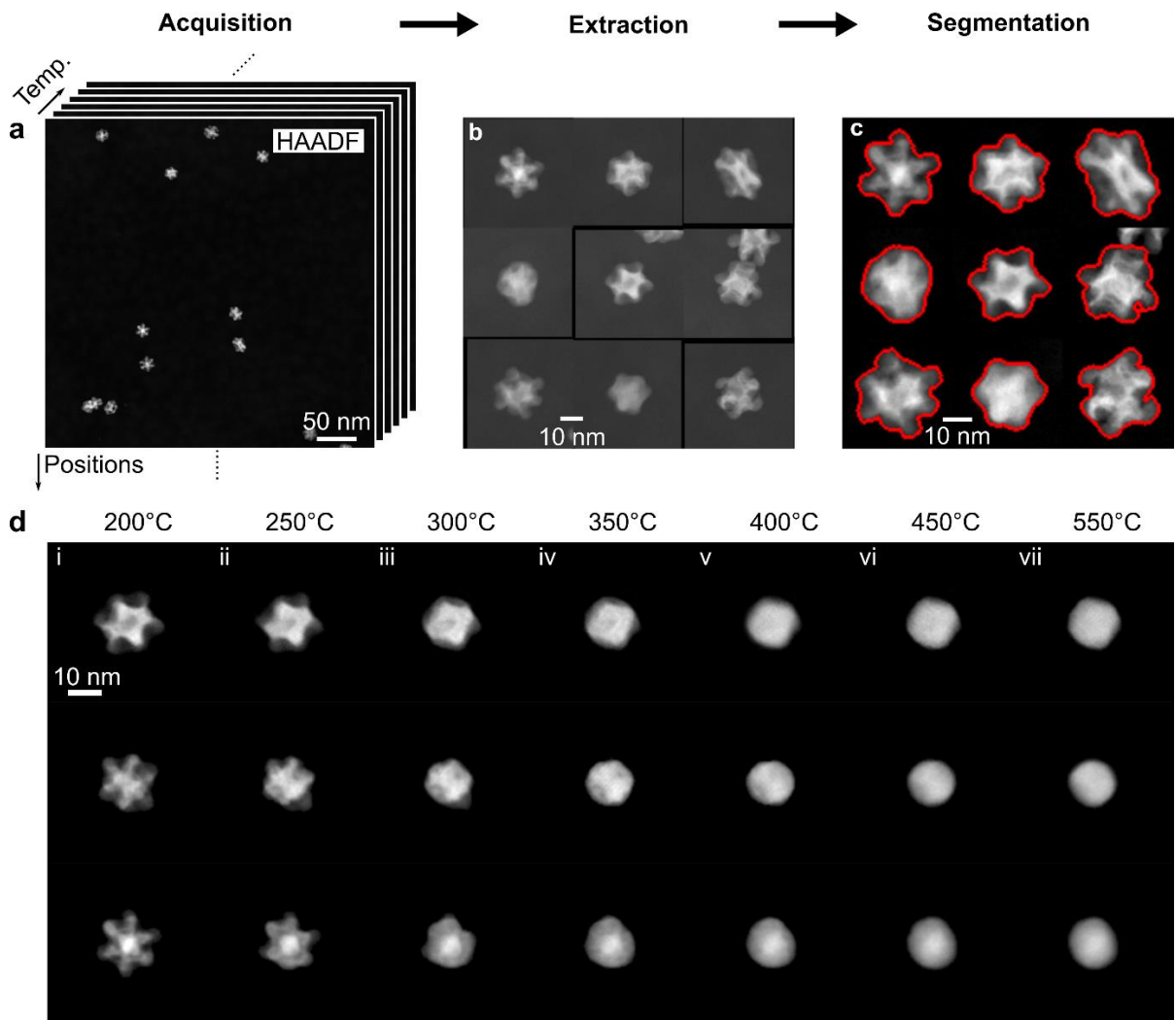

**Figure 1. Overview of methodology applied to extract 2D projections of PtNi nanoparticles as a function of temperature.** (a) As-acquired STEM-HAADF images stacked as multidimensional data sets, with dimensions of temperature and x,y location on the support grid. (b) Individual nanoparticles are extracted into small square regions. (c) The boundaries of the nanoparticles are identified and outlined in red using watershed segmentation. (d) STEM-HAADF images showing three examples of extracted nanoparticles as a function of temperature at (i) 200°C (ii) 250°C, (iii) 300°C, (iv) 350°C, (v) 400°C, (vi) 450°C, and (vii) 550°C.

The SPR methodology requires that all particles are sufficiently similar and, therefore, quantification of variations in morphology and composition for the nanoparticles at the different temperatures used in this study was performed (Figure 2). The STEM-HAADF signals were used to characterize the nanoparticles' morphology and the STEM-EDX signals were used to quantify nanoparticle composition. The normalized area change of each individual nanoparticle was plotted as a function of temperature to reveal the mean transformation behavior and the variability in the behavior of individual particles upon heating (Figure 2a). The majority of nanoparticles were found to transform in a similar manner, with the standard deviation less than ±10% (blue line in Fig. 2a), validating the use of SPR for mean 3D structure reconstruction in this PtNi system. It is important to note that 2D area data contains inherent variability due to differences in the projected cross-section under different orientations, so is not solely due to nanoparticle dissimilarity (exemplified in Figure 1d, where at 200°C the three nanoparticles have similar morphology but different projected areas). The particles show a rapid loss of projected area (approximately 20% reduction) on heating from 200°C to 350°C, followed by a slower change from 350°C to 550°C (approximately 1% reduction) (Figure 2a). The trend of projected area change is the same for both small and large particles when finer categorization is performed on the 60 nanoparticles according to their mean projected area at 200°C (SI Figure S3).

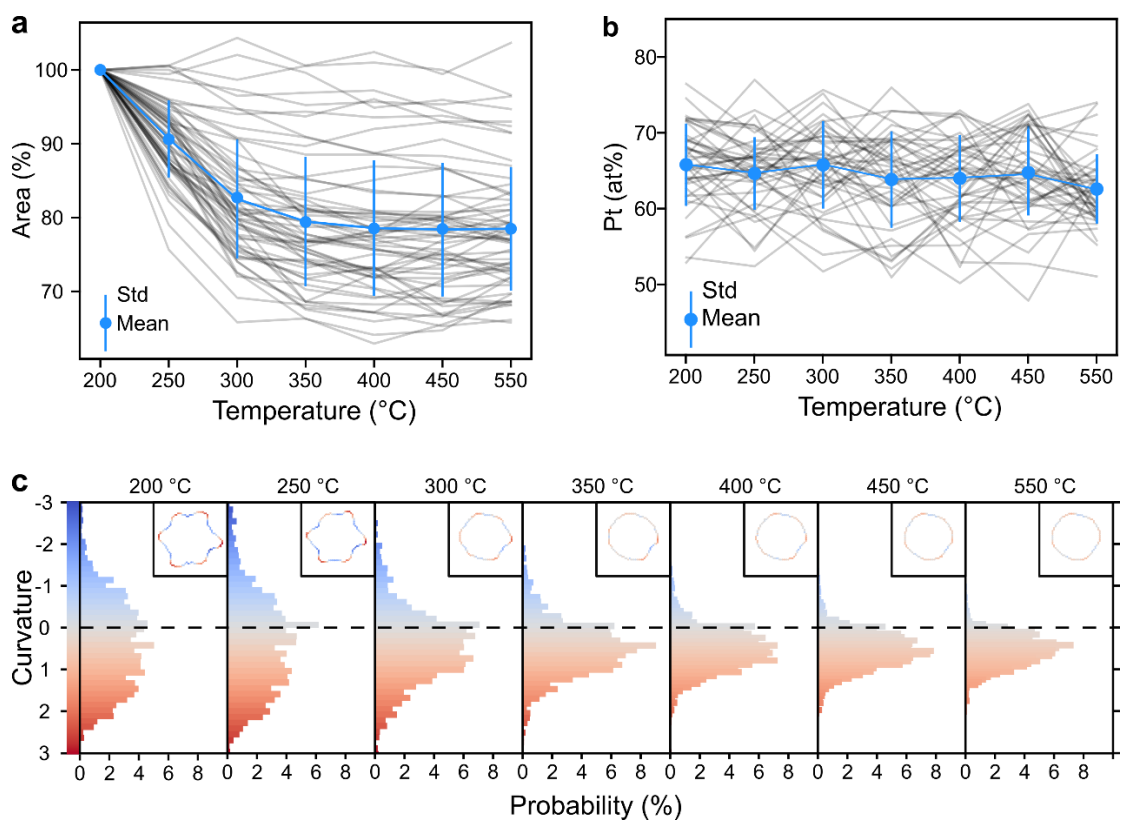

**Figure 2. 2D projection analysis of an ensemble of 60 nanoparticles.** (a) Normalized 2D projection area of 60 individual nanoparticles (grey lines) and averaged particle area (blue line) as a function of heating temperature. (b) Pt atomic percentage of 60 individual nanoparticles (grey lines) and averaged Pt at% (blue line) as a function of heating temperature. (c) Normalized curvature of 60 individual nanoparticles as a function of heating temperature. A curvature of 0 represents a flat surface while that of 1 (-1) represents a convex (concave) surface fitted with a circle having a radius equal to the equivalent radius of the nanoparticle. Insets are projected boundary curvature changes of an example nanoparticle (HAADF images shown in Figure. 1d top row) as a function of temperature.

Examining the STEM-HAADF images of three example particles in Figure 1d reveals the restructuring process of the PtNi rhombic dodecahedron nanoparticles; transforming from a complex structure possessing concave surfaces to a near spherical morphology. To further analyze such morphological deformation, curvature changes of all 60 nanoparticles were tracked (Figure 2c) based on their STEM-HAADF projections. The curvature is defined as 1/r, where r is the radius of a fitted circle for the boundary segment, which is normalized by dividing the equivalent radius of the nanoparticle to show either a concave (curvature < 0) or convex surface (curvature > 0). In general, the curvature distributions are broad at early stages of the heating, demonstrating that the nanoparticles consist of concave and convex surfaces. At higher temperatures, specifically 350°C to 550°C, the curvature distribution narrows with a maximum around 0.5, demonstrating the nanoparticles overall transformation to an approximately spherical shape with positive curvature. This dramatic change in morphology is accompanied by very little compositional change (remains as 65±5 Pt at%) at each temperature step, suggesting no leaching of either Pt or Ni (Figure 2b).

**3D structural transformation of PtNi rhombic dodecahedra during heating**

The 2D projection analysis in Figure 2 can fail to detect information from the third dimension, so only provides a limited understanding of the structural and compositional transformation. In-situ SPR is therefore applied, exploiting the mass-thickness dependence of the STEM-HAADF signal, to perform 3D reconstruction of nanoparticle morphology and obtain qualitative information on elemental distribution. To perform a 3D reconstruction using SPR, the orientation of the individual as-segmented nanoparticles needs to be determined. We used a template matching approach to match experimental in-situ images with re-projections of known orientation, generated from our ex-situ 3D reconstruction of the same nanoparticles at room temperature (SI section 1. 2).[27] As most of the nanoparticles lose their well-defined shapes during heating, orientation matching at elevated temperature is inaccurate. Therefore, we assume that nanoparticles have not undergone a substantial orientation change during the heating process and use the morphology of nanoparticles at 200°C to assign the orientation throughout the temperature series. To validate this assumption, one nanoparticle is imaged with atomic resolution from 200°C to 500°C (SI Figure S4). The atomic resolution HAADF images and accompanying fast Fourier transforms (FFTs) show a similar atomic configuration was maintained through heating, verifying the assumption that the nanoparticle does not undergo significant rotation during heating.

In the 3D reconstructions (Figure 3), the nanoparticles are found to have a concave rhombic dodecahedral morphology at 200°C (Figure 3a and c i), which is in good agreement with the observed room temperature structure.[27] Upon heating beyond 200°C, the most prominent feature observed is that the highly concave {110} facets (Figure 3 i) become less concave from 200°C to 250°C (Figure 3 ii), then gradually change to flat surfaces from 300°C to 350°C (Figure 3 iii and iv) in agreement with the 2D curvature analysis. At 350°C, the nanoparticle remains a rhombic dodecahedron, although shows no concave surface features. From 350°C to 550°C, the surface features on the particles gradually become less distinct until the morphology is approximately spherical (Figure 3 iv-vii and b). This surface evolution broadly agrees with previous 2D S/TEM reports on concave octahedral PtNi nanoparticles.[8,30] However, this 3D reconstruction unambiguously distinguishes that the surface evolution is dominated by flattening of concave surfaces, and is based on a larger number of particles than have been considered previously.

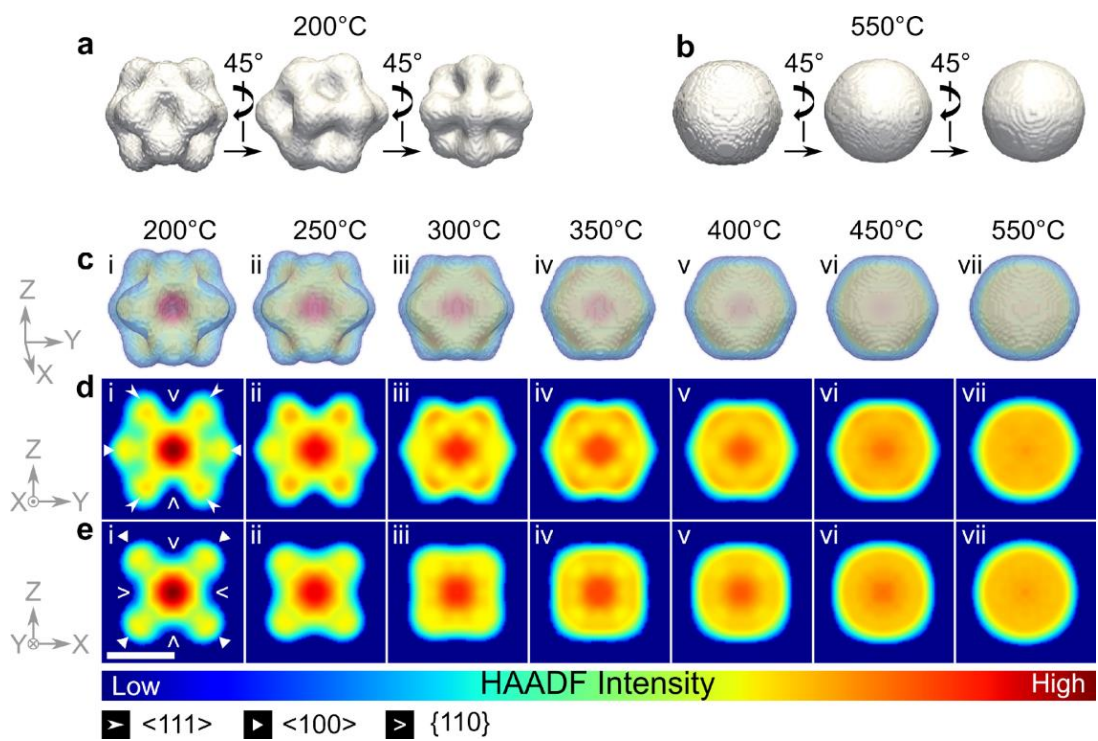

**Fig 3. In-situ single particle reconstruction STEM-HAADF results revealing the 3D structural evolution of PtNi nanoparticles during heating.** Opaque surface render views at various orientations revealing nanoparticles' 3D morphology at (a) 200°C and (b) 550°C. (c) 3D volume intensities revealing nanoparticles' 3D morphology at each temperature (transparent surface render). (d, e) Orthoslices extracted from the center of the 3D volume intensities (c), parallel to Y-Z (d) and X-Z (e) planes. Higher STEM-HAADF intensity in the core reflects a higher Pt concentration as a result of the higher atomic number of Pt versus Ni. In (i), pointed arrows indicate <111> vertices, open V arrows indicate {110} surfaces and closed triangles indicate <100> vertices. Scale bar is 10 nm and applies to all panels.

The atomic number sensitivity of the STEM-HAADF imaging mode, coupled with our 3D SPR tomographic data allows us to go beyond surface morphology and investigate changes in the nanoparticles' internal compositional distribution. Internal elemental redistributions (Figure 3c and d) are revealed by false colored orthoslices extracted from the 3D data. In this analysis the mean nanoparticle composition is assumed to be constant during heating, which is consistent with our STEM-EDX data, within measurement accuracy (Figure 2b). As the particles are single crystals, high intensity regions indicate Pt enrichment due to its higher atomic number compared to Ni. Analysis of the 3D intensity difference between consecutive temperature steps is shown in SI Figure S5. The internal compositional data shows that the high intensity core continuously reduces in intensity (Figure 3 i-vi), with the particle finally reaching uniform intensity, corresponding to a uniform internal composition, at 550°C (Figure 3 vii). The high intensity core seen at the start of the heat treatment is attributed to a Pt-enriched core inherited from the Pt seeds used to nucleate the particles.[31,32] The reduction in STEM-HAADF intensity in the center of the particle therefore demonstrates the rate that the Pt from the Pt-seed gradually diffuses through the structure as the nanoparticle composition tends towards homogeneous alloying.

A more detailed study of the intensity distributions in Figure 3c and e reveals that high intensity Pt-enriched regions are also observed on the <111> vertices at the start of heat treatment (positions indicated by pointed arrows in Figure 3 i), in agreement with the previous room temperature elemental SPR result.[27] This enrichment increases relative to the core intensity from 200°C to 250°C, Figure 3c i-ii, indicating preferential diffusion of Pt to <111> vertices (e.g. the lowest surface energy facets)[33] and that surface Pt enrichment is not lost despite homogenization of the core. The preferential diffusion of Pt may be driven by reducing lattice strain within the particle: the larger lattice parameter of Pt is known to be more favorable than Ni for undercoordinated surfaces.[34] From 250°C to 300°C, the higher intensity (high Pt concentration) regions at the <111> vertices become less distinct as the particle becomes less concave (Figure 3c ii and iii). From 300°C to 350°C, the intensity of <111> vertices reduces (Figure 3c iv) and a higher intensity layer is visible on {110} facets (Figure 3d iv, {110} facets are indicated by open V arrows in Figure 3c i). At 350°C, the nanoparticle composition can be roughly considered as a core-shell-shell structure, with Pt enriched at the outer surface and inside the core (Figure S6). From 350°C to 550°C, the intensity differences reduce, reaching a homogeneous composition distribution for the whole nanoparticle.

**Atomic structure transformation of PtNi rhombic dodecahedra during heating**

The voxel resolution of the 3D SPR reconstruction is at the nanoscale, so atomic scale structural and compositional transformations are not resolved. To assess the extent of atomic restructuring, we therefore obtained high resolution STEM-HAADF images of a <110> oriented nanoparticle at room temperature and after various heating times from 5 seconds to 300 seconds at 400°C (Figure 4).

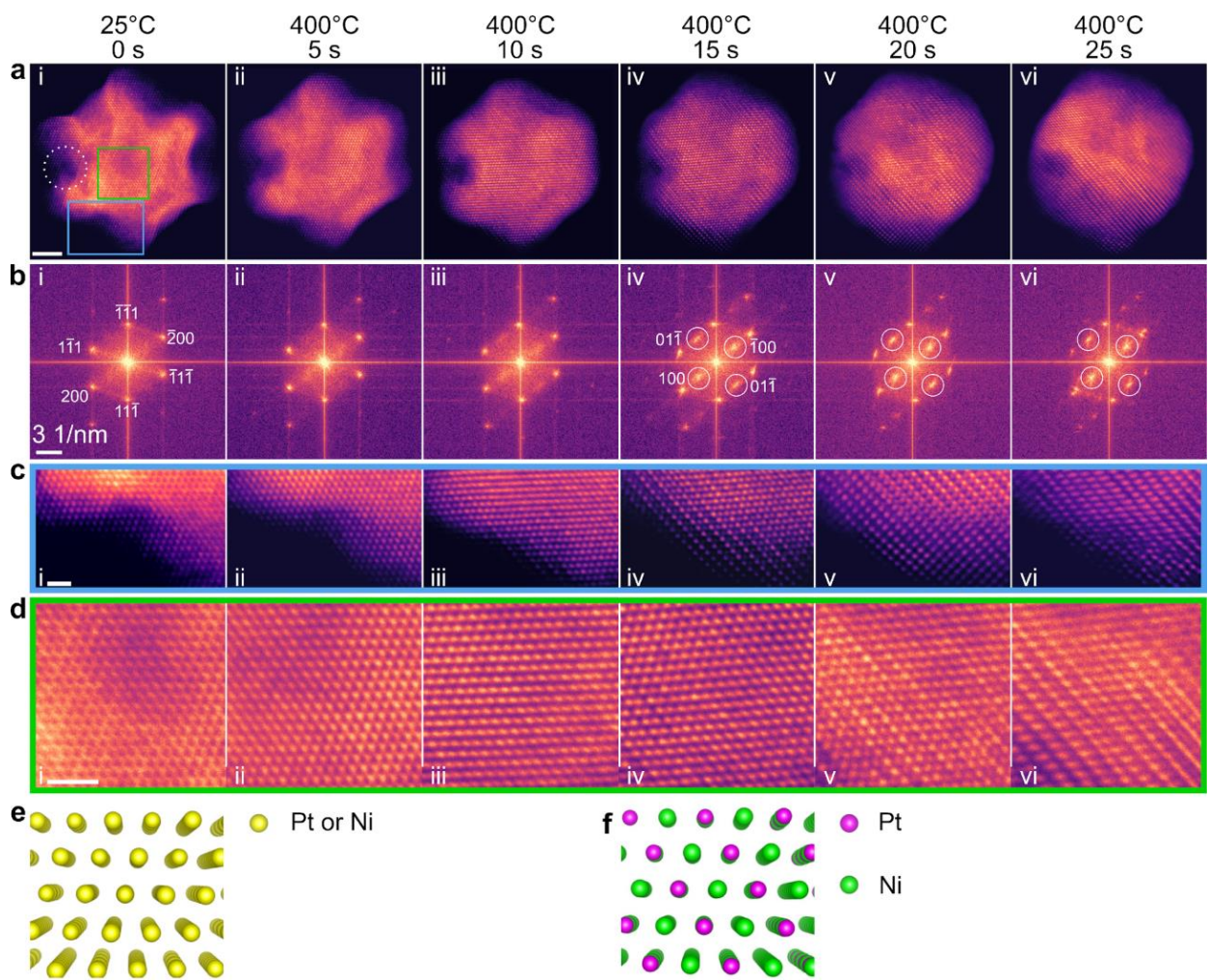

**Figure 4. Atomic structure reconfiguration of a PtNi nanoparticle during heating.** (a, b) Atomic resolution HAADF images (a) and corresponding FFTs (b) of the nanoparticle at (i) room temperature (25°C) and (ii-vi) after heating at 400°C for 5 s, 10 s, 15 s, 20 s, 25 s, respectively. (c, d) Enlarged views of an outer and an inner particle region, respectively. (e, f) atomic model demonstrating the disordered A0 phase and chemically ordered L1$_2$ phase, respectively. Scale bars in (a), (c), and (d) are 3 nm, 1 nm, 1nm, respectively.

Observation of the PtNi nanoparticles at atomic-resolution after heat treatment at elevated temperatures revealed a transformation from a disordered alloy to an ordered lattice. At room temperature and for short heating times (5 and 10 seconds) at 400°C (Figure 4i-iii), all nearest neighboring atomic column intensities are absent of any periodic variation, the presence of which indicates the L12 ordered phase. Smaller, aperiodic differences in the intensity of neighboring atomic columns can be assigned to inhomogeneous alloying at the atomic scale as well as to surface steps and other defects, so the observed atomic column intensities suggest that the Pt and Ni atoms are randomly arranged in the face-centered cubic (FCC) unit cell as a disordered alloy (A0 phase, Figure 4e). . In contrast, after 15 seconds heating at 400°C, atomic columns are rearranged into regular dark and bright lines, indicating the formation of an ordered $L1_2$ phase in these surface regions (Figure 4a-c iv-vi and f). The formation of the ordered phase is also evidenced by the emerging reflections in the Fourier transforms, as circled in Figure 4b iv-vi. The ordered phase is first visible in the region of the <111> vertices on the imaged particle (Figure 4c iv) after 15 seconds heating, but emerges in the inner region slightly later (after 20 seconds heating, Figure 4d v). We suggest this difference is related to the presence of the surface, increasing the diffusion rate locally and accommodating strains, thus facilitating formation of the ordered $L1_2$ phase. On further heating for up to 180 seconds the ordered phase extends over most of the nanoparticle, although the precise orientation and degree of ordering varies at the nanoscale (Figure 4 iv-vi and Figure S7). Heating for longer times (up to 300 seconds) did not significantly alter the nanoparticles' visible lattice structure (Figure S7), suggesting the nanoparticle ordering is complete at 180 s.

Chemical ordering is known to occur in PtNi and other Pt based nanoparticles under heating.[35,36] Ordered phases have been reported to show higher catalytic performance for the oxygen reduction reaction and higher structure stability but a poorer performance in the hydrogen evolution reaction.[37] It is believed that heating facilitates the formation of the thermodynamically favored ordered PtNi phase through increases in the rate of atomic diffusion and by favoring a larger number of defect sites, facilitating the formation of the thermodynamically favored ordered phase.[38] In the particular case of the PtNi particles here, we also postulate that the homogenization of composition contributes towards the formation of the ordered $L1_2$ phase, where the homogenized composition of 65 at% Pt is within the composition range of the formation of a $Pt_3Ni_1$ $L1_2$ phase.[39,40] Prior to heating, the variations in composition (particularly outside the range in which the ordered phase forms) should prevent the formation of the ordered phase across the particle as a whole.

**Catalytic performance of the PtNi electrocatalysts**

To better understand the effect of the structural and compositional changes induced by post-synthesis thermal treatment on the electrocatalytic activity of nanoparticles, we have also tested the electrochemical performance of these nanoparticles with and without heat treatment. For electrocatalytic testing the PtNi colloids were mixed with a highly conductive and high surface area carbon (C) support, yielding a metal loading which approximated 20 wt.% (see experimental methods for details). The as-synthesized supported PtNi nanoparticles (hereafter referred to as PtNi/C) were compared to those after heat-treatment at 400ºC for 12 hours (PtNi/C-heated). The size distribution of PtNi/C is narrowly peaked at 11.6 nm, while PtNi/C-heated is more widely distributed and peaked at 10.3 nm (Figure 5 a-c), corresponding to an approximately 10% size reduction in the PtNi/C-heated sample. However, the 2D projection images of the PtNi/C-heated sample did not show any large differences in morphology compared to the starting material and no atomic ordering was observed (SI Figure S9). Surprisingly, this 10% shrinkage and nanoparticle morphology after heating the PtNi/C at 400ºC corresponds to the equivalent size reduction and shapes observed during heating at just 250ºC in the in-situ SPR experiment. This may suggest that the transformation rate for supported nanoparticles is strongly affected by the type of substrate. The PtNi nanoparticles are found to be more stable on the carbon support compared to when heated on the SiN chips, which could be explained by the high lateral and low inter-layer thermal conductivity of the carbon support[41] dissipating external heat, or the support-metal interaction on carbon stabilizing the nanoparticles[42], as well as the overlayed carbon layer inhibiting nanoparticles' surface atom diffusion.

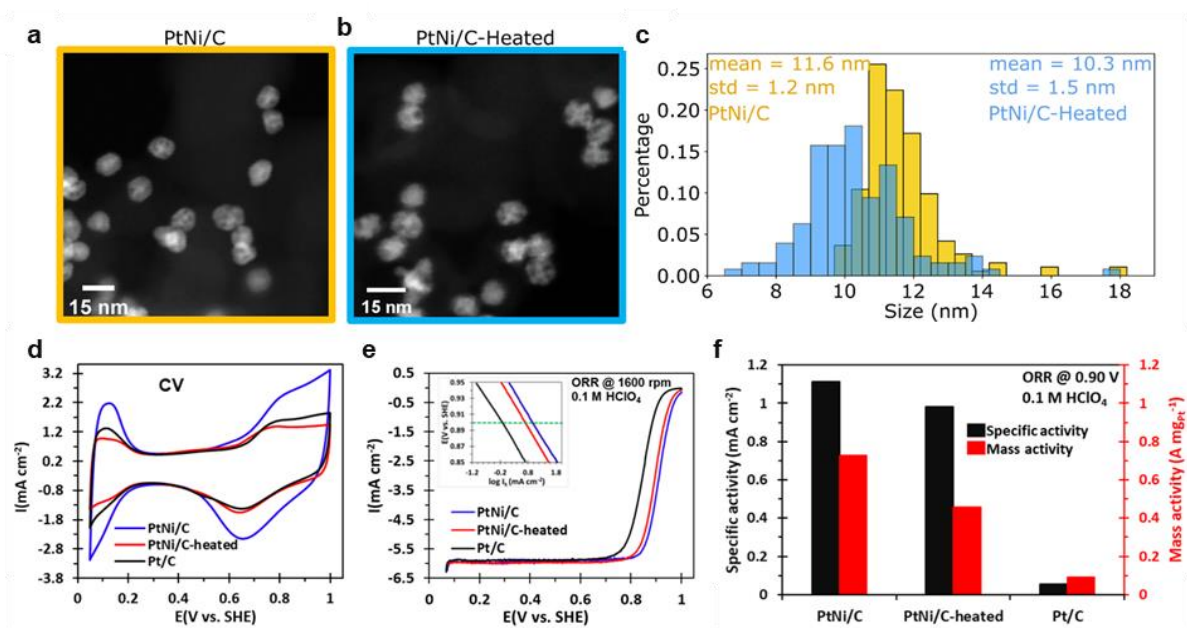

**Figure 5. Catalytic performance tests of PtNi nanoparticles.** a, STEM-HAADF images of PtNi/C nanoparticles. b, STEM-HAADF images of PtNi/C-heated nanoparticles. c, Size distribution histograms of PtNi/C (yellow) and PtNi/C-heated (blue) samples. d, Cyclic voltammograms of PtNi/C (blue), PtNi/C-heated (red) and Pt/C (black) electrocatalysts recorded in Ar-saturated 0.1 M HClO₄ electrolyte solution at room temperature with a scan rate of 100 mV s⁻¹ after 30 potential cycles of catalyst activation. e, Background subtracted and *iR* corrected ORR polarization profiles recorded in O₂-saturated 0.1 M HClO₄ solution with a sweep rate of 10 mV

s$^{-1}$ at 1600 rpm and normalized by the glassy carbon area (0.196 cm$^2$). The ORR Tafel plots are the inset in (e). f, Intrinsic area-specific activities and mass-specific activities of PtNi/C and PtNi/C-heated at 0.90 V (*vs* SHE), compared with the Pt/C electrocatalyst.

To evaluate the nanoparticles' performance for the ORR, we used cyclic voltammograms (CVs) (Figure 5d) and carbon monoxide (CO) stripping profiles (SI Figure S10a) to estimate the electrochemically active surface areas (ECSAs) of these PtNi/C catalysts (~20 wt.% loading) and compared these to commercial state-of-the-art Pt/C (Alfa Aesar, HiSpec, 20 wt.% metal loading) electrocatalysts. Both ECSA$_{Hupd}$ (H$_{upd}$ = underpotential deposited hydrogen) and ECSA$_{CO}$ were calculated by normalizing the measured charges Q$_H$ of hydrogen adsorbed (H$_{upd}$, using 210 μC/cm$^2$$_{Pt}$)[43–46] and CO adsorbed (CO$_{ads}$, using 420 μC/cm$^2$$_{Pt}$)[47] on the electrocatalysts. The corresponding specific ECSA$_{Hupd}$ and ECSA$_{CO}$ plots are shown in Figure S10b. The unusually high ECSA$_{CO}$ values originated from the double responses in CO current peaks on the PtNi/C electrocatalysts, resulting in current density peak broadening. We ascribe the first peak (lower potential) to CO adsorption on PtNi nanoparticles with near surface rich-Ni whereas the second current peak (higher potential) is associated with Pt-rich surfaces (low Ni content). This second peak is less pronounced after the heat treatment, suggesting that the surface is more uniformly alloyed (consistent with the behavior found from the in-situ SPR data). Furthermore, all the CO-stripping oxidation peaks for the alloy PtNi/C and PtNi/C-heated electrocatalysts (0.55–0.75 V, *vs* standard hydrogen electrode (SHE)) show a negative potential shift of ~132 mV relative to that of the Pt/C electrocatalyst (0.60–0.85 V, *vs* SHE), suggesting enhanced CO-poisoning resistance. Background subtracted and solution resistance (*iR*) corrected ORR polarization curves recorded in O$_2$-saturated 0.1 M HClO$_4$ solution with a sweep rate of 10 mV s$^{-1}$ at 1600 rpm are compared in Figure 5e (inset is the corresponding Tafel plots). The polarization profiles show two distinct regimes: a diffusion-limited current density region (~0.05–0.85 V, $vs$ SHE) and a mixed kinetic-diffusion-controlled regime (the true measure of the catalyst functionality) (0.85–1.00 V, $vs$ SHE). The ORR polarization curves for all the electrocatalysts reached the diffusion limited-current density at ~6.0 mA cm$^{-2}$ (geometric), consistent with the reported theoretical values (~5.8–6.02 mA cm$^{-2}$).[48–50] The positive shifts to higher potentials for the PtNi/C and PtNi/C-heated electrocatalysts demonstrate enhanced kinetics for the electrocatalytic ORR performance relative to the Pt/C catalyst. The measured currents from ORR polarization curves were corrected for mass transport to acquire the true kinetic currents. The kinetic currents were calculated using the Koutecky–Levich equation.[49]

To compare the nanoparticles' ORR catalytic performance, the specific activity and mass activity were obtained by normalizing the kinetic currents with ECSA$_{Hupd}$ and working electrode (WE) Pt mass loading, respectively. The area-specific activities (Figure 5f) calculated at 0.90 V (vs SHE) are 1.11 mA cm$^{-2}$ (PtNi/C), 0.98 mA cm$^{-2}$ (PtNi/C-heated) and 0.09 mA cm$^{-2}$ (Pt/C). These area-specific values represent improvement factors of 12.3 (PtNi/C) and 10.9 (PtNi/C-heated) relative to the Pt/C catalyst. Based on Pt mass loading, the mass-specific activities (Figure 5f) were calculated to be 0.73 A mg$_{Pt}^{-1}$ (PtNi/C), 0.46 A mg$_{Pt}^{-1}$ (PtNi/C-heated) and 0.06 A mg$_{Pt}^{-1}$ (Pt/C). Thus, the PtNi/C and PtNi/C-heated electrocatalysts achieved 12.2- and 7.8-fold enhancement when compared to the performance value of the Pt/C catalyst. The electrocatalytic ORR performance enhancement of the PtNi/C electrocatalyst is ascribed to its well-defined concave morphology, whereas the catalytic decline of the PtNi/C-heated electrocatalyst is associated with the loss of initial concave facets during post-synthesis heat-treatment. These findings are consistent with in-situ SPR investigations and STEM imaging, in that fewer concave surface sites (high-coordination number) are available after heating. The decrease in specific activity will also be associated with the loss of Pt-rich surface regions as the nanoparticles become more homogeneously alloyed. These effects should run counter to the effect of the transition to an ordered structure, which has previously been shown to have a positive effect on the ORR activity of Pt-based nanoparticles.

**Potential of the SPR approach for in situ TEM studies**

This proof of principle investigation demonstrates the potential of the in-situ 3D SPR imaging approach for nanoscale 3D in-situ investigations. Compared to traditional tilt-series TEM tomography, the SPR method has the major advantage that specimen tilting it is not required to obtain a 3D reconstruction. This allows 3D imaging of nanoparticle samples even when using bulky in-situ holders, whether that is the heating holder demonstrated in this work or in-situ gas and liquid holders, which usually cannot be tilted beyond 30° for high resolution TEM instruments due to the small polepiece gap. The SPR technique therefore opens the field to new possibilities for 3D studies of nanoparticle evolution in gaseous and liquid environments. Removing the need for specimen tilting also allows better time-resolution for in-situ studies compared to tilt-series TEM tomography, where the necessity to take 10s of images for each 3D reconstruction generally limits the achievable time resolution to hours or days. For our SPR approach one image is enough if there are sufficient particles in the field of view, and temporal resolutions of the order of seconds could be achievable.

Another big advantage of SPR is that it does not provide information for a "single nanoparticle" but uses an ensemble of nanoparticles (60 nanoparticles were used in this study) to reconstruct an averaged 3D representation of the changes in the population during the in situ experiment. This has the significant benefit of lowering the necessary electron dose imposed on each nanoparticle, as each particle is only imaged once for a single 3D reconstruction. Multiple tilt series tomographic data sets quickly increase the total electron dose required, often preventing in-situ tilt series tomographic reconstruction for electron beam-sensitive nanoparticles, like the PtNi studied here. The required electron dose can be lowered further by increasing the number of particles in the ensemble averaging, therefore, the SPR method provides new opportunities for in situ characterization of beam-sensitive specimens. As a consequence of averaging nanoscale information over many nanoparticles, the SPR results are more representative of the nanocatalysts population, reducing the chance of outliers being used for interpretation of structure-property relationships. In comparison, the tilt-series tomography method often uses only one nanoparticle to represent large populations.

The major current disadvantage of the SPR technique is the need for nanoparticles of near-identical structure to be included within the reconstruction, since averaging particles with local structural or compositional differences will limit the spatial resolution of the achievable reconstruction. In this work the limited spatial resolution of the 3D reconstruction prevents the atomic ordering within the $L1_2$ phase being resolved and we have used atomic resolution 2D images to observe this, complementing the nanoscale 3D reconstruction. Nonetheless, there is no inherent spatial resolution limit for SPR, and achieving atomic resolution only requires more similar particles be included in the reconstruction and/or greater sampling of each particle. We have recently shown that SPR is achievable even for heterogeneous nanoparticle systems by acquiring a larger number of images and using advanced algorithms to separate disparate particle populations, producing separate reconstructions for different nanoparticle classes [ref 51] although this study used a model nanoparticle system. Higher spatial resolution could also be achieved by using a higher electron flux for more stable nanoparticles compositions, by considering smaller, or less compositionally complex nanoparticles, magic number nanoparticles [ref 52] or by exploiting crystal symmetry to combine information from symmetrically equivalent features, as is sometimes done in biological SPR, for example to enable atomic resolution reconstruction of catalytically active surface sites at particular apices.

The other disadvantage of the inorganic SPR approach is that the data processing steps are still quite manual and laborious. There is not yet a routine start-to-finish software approach that can be applied to reconstruct an arbitrary inorganic nanoparticle ensemble data set. Nonetheless, automated image analysis in EM is attracting significant research interest and we believe that fully automated processing could be achievable in the future.

**CONCLUSION**

In this work, we have extended the single particle reconstruction method from ex-situ room temperature analysis conditions to in-situ analysis of temperature induced structural changes. We study the evolution of PtNi rhombic dodecahedral nanoparticles with temperature, revealing significant changes in the average nanoparticle 3D structure and compositional distribution at each temperature step using STEM-HAADF signals. During heating from 200°C to 550°C, the Pt from the seed at the center of the PtNi rhombic dodecahedral nanoparticle diffuses to the surface, resulting in a lower Pt concentration in the core but increased surface enrichment, especially at <111> vertices. Concave surfaces are more easily deformed than convex surfaces during heating, with the nanoparticles finally reaching a thermodynamically favored spherical shape with a homogenous elemental distribution. The specific activity in the ORR for the synthesized PtNi nanoparticles declines by 50% when the catalysts are heated at 400°C, demonstrating the advantage of the initial concave surface morphology. Our SPR 3D reconstruction approach provides new evidence of the small morphological transformations and compositional changes that cause a decline in catalytic performance for the ORR, although further work is needed to understand the effect of the catalytic support on controlling the rate of this behavior.

In addition, this proof-of-concept study demonstrates the potential of the SPR approach for in situ time resolved 3D nanoscale studies without the need for sample tilt. The approach provides new opportunities for complex experiments with electron beam sensitive samples and when using liquid and gas-cell environmental holders. We envisage that the in-situ 3D imaging methodology could also be combined with automated acquisition schemes to generate high throughput 3D data.[51] In situ SPR imaging therefore opens up the possibility to study the 3D in-situ shape and chemical transformation of nanoparticles under heating, electric, gaseous, liquid, light or other stimuli. This would be very valuable for understanding structure-property relationships of large populations of inorganic nanoparticles and hence accelerate their use in applications such as energy harvesting, catalysis and medical imaging.

**METHODS**

***Nanoparticle synthesis :*** PtNi rhombic dodecahedral nanoparticles were synthesized through a standard co-reduction between nickel acetate tetrahydrate and chloroplatinic acid solution in the presence of ternary hydrophobic surfactants oleylamine (OAm), octadecylamine (ODA) and oleic acid (OLEA), as detailed in our previous work.[5] The as-synthesized nanoparticles were re-suspended in chloroform, followed by 5 minutes sonication.

***TEM sample preparation for in-situ imaging:*** A Protochips Fusion 500 double tilt heating holder and MEMS based heating chips (series number E-FHBS, covered by a 40 nm thick continuous SiN support film) were used for deposition of the nanoparticle solution and to perform heating of the specimen inside the microscope.

To minimize the contamination during long STEM acquisition times, heating chips were vacuum baked for 2 hours at 100°C and plasma cleaned for 5 minutes before the deposition of specimen. After drop-casting the specimen on heating chips, a few drops of methanol were also drop-cast onto the chip to increase nanoparticle dispersion and to prevent severe agglomeration of the nanoparticles. Heating chips were then vacuum baked for 12 hours at 100°C and plasma cleaned for 15 seconds, before loading onto the heating holder. The holder was loaded into the microscope 12 hours before the start of the experiment to further reduce contamination in the microscope. Before data acquisition, a 30-minute beam shower was performed on the viewing region.

***STEM and EDX data acquisition:*** A Thermo Fisher Titan G2 80-200 S/TEM was used for STEM-HAADF and EDX spectrum image data acquisition. The microscope was operated at 200 kV, with a beam current of 180 pA, equipped with an X-FEG high brightness source and STEM probe aberration corrector. All HAADF and EDX data were acquired at 225,000X magnification with a full frame size of 1024 x 1024 pixels, resulting in a pixel size of 3.86 Å and a field of view of 395 nm by 395 nm. For STEM-HAADF imaging, a 21 mrad convergence angle and 55 mrad acceptance inner angle were used. STEM-HAADF images were acquired in the Thermo Fisher TIA software using a pixel dwell time of 20 μs. For EDX spectrum imaging, the Bruker Esprit software was used for data acquisition using a Super-X EDX detector system consisting of four silicon drift detectors (SDDs) with a total collection solid angle of approximately 0.7 sr. A pixel dwell time of 50 μs and 15 scanned frames were summed to produce each EDX spectrum image data set, resulting in a total pixel dwell time of 750 us. The EDX spectrometer was set to collect an energy range from 0-20 keV with 2048 channels.

At room temperature, six regions of interest (ROI) were selected for characterization during heating. The selection criterion for the ROI was to ensure as many nanoparticles were included in the field of view as possible but without the individual particles overlapping. This ensured that the later particle extraction image processing steps can be accurately performed and that the nanoparticles do not melt into adjacent nanoparticles during heating. Each ROI contained about 10 well dispersed nanoparticles. The heat treatment was performed at 200°C, 250°C, 300°C, 350°C, 500°C, 450°C and 550°C with a heating ramp of 10°C s$^{-1}$ and a total heating time of 10 minutes at each temperature step. After heat treatment, the specimen was cooled to room temperature (cooling rate 100°C s$^{-1}$), then the 6 selected regions were characterized by STEM-HAADF and STEM-EDX spectrum imaging. For atomic resolution imaging, the specimen was heated at 400°C for 5s, 10s, 15s, 20s, 25s, 60s, 180s, and 300s. After each heating time, the specimen was cooled to room temperature (cooling rate 100°C s$^{-1}$) for STEM-HAADF image acquisition.

***Preparation of carbon-supported electrocatalysts:*** Practical electrocatalysts were created by dispersing the as-synthesized nanoparticles onto a carbon support (Cabot, Vulcan XC-72R) via a colloidal-deposition strategy, by mixing these alloy colloids in chloroform with carbon black, followed by vigorous sonication for 10 minutes.[5, 53] The resulting homogeneous dark brown mixture was left in a fume hood overnight to evaporate the chloroform. The resultant carbon-supported materials were further washed with acetone 5 times to eliminate residual surfactants/impurities and dried overnight in an oven at 60ºC. The catalyst was subjected to a 400ºC heat treatment for 12 hours in a furnace.

***Structural characterization of carbon-anchored alloy electrocatalysts:*** Specimens for scanning transmission electron microscopy (STEM) investigations were prepared by one-drop casting of a suspension of these electrocatalysts in acetone onto carbon-support films on copper grids and allowed to air dry under ambient conditions. Specimens were analyzed using HRTEM and STEM-HAADF on a ThermoFisher Tecnai F20 FEG TEM operating at 200 kV.

***Electrochemical Measurements:*** Electrocatalyst inks were prepared by mixing 10 mg of the electrocatalyst with 2 ml of Milli-Q water (Millipore, 18.2 M$\Omega$.cm @ 25°C), 0.4–0.5 ml isopropanol and 25 µl Nafion® perfluorinated resin solution (5 wt.% in a mixture of lower aliphatic alcohols and water (45% water)). The resultant mixture was sonicated for 15 minutes, yielding a dark-brown homogeneous catalyst ink. A 10 µl of the catalyst ink was then carefully pipetted onto a polished glassy carbon (GC) electrode (Pine Research Instrumentation, 5 mm disk OD) and allowed to dry under ambient conditions. The remaining thin black uniform film of Nafion-electrocatalyst-carbon across the GC served as the working electrode (WE). All the experiments were performed using a Biologic SP300 potentiostat coupled with Gamry instrument RDE710 Rotator at room temperature. A Pt wire and mercury/mercurous sulphate served as the counter and reference electrodes, respectively. All potential values were reported relative to the standard hydrogen electrode (SHE). The readout currents were corrected for the Ohmic iR losses.

Cyclic voltammetry (CV) curves were recorded by subjecting the WEs in Ar-purged 0.1 M HClO$_4$ electrolyte solution to pre-activate the electrocatalysts for 30 potential cycles between 0.05 V and 1.10 V (*vs* SHE) at the scan rate of 100 mV s$^{-1}$. The sweep rate was then reduced to 50 mV

s$^{-1}$, the second CV cycles were used for analysis. The electrochemically active surface area (ECSA) was calculated by integrating the area under the curve for the hydrogen underpotential deposition region (H$_{upd}$), assuming a monolayer hydrogen charge of 210 µC/cm$^2_{Pt}$.[54]    Carbon monoxide (CO) stripping voltammetry was acquired by bubbling CO gas into the electrolyte solution while holding the potential of the working electrode at 0.1 V (*vs* SHE). Dissolved or residual CO gas was removed from the electrolyte by purging with Ar while still holding the potential of the WE at 0.10 V (*vs* SHE). The potential of the WE was then cycled to 1.00 V (*vs* SHE) at 20 mV s$^{-1}$, followed by a CV cycle as described above at 20 mV s$^{-1}$. The peak area could then be determined using the baseline CV and a normalization factor of 420 µC/cm$^2_{Pt}$[47] was used to calculate the ECSA.

ORR polarization curves were recorded at a rotation speed of 1600 rpm in an oxygen (O$_2$)-saturated 0.1 M HClO$_4$ electrolyte and corrected for the capacitive current associated with Pt-solute/C electrocatalysts by subtracting a CV curve measured in Ar-purged 0.1 M HClO$_4$. The current densities were also normalized with reference to the calculated ECSA to evaluate the specific activities.

ASSOCIATED CONTENT

**Supporting Information**.

The following files are available free of charge.

Image processing, consideration of microscope acquisition parameters for in-situ SPR, area identification error analysis and additional experimental results (PDF).

AUTHOR INFORMATION

**Corresponding Author**

*Thomas J. A. Slater, Email: slatert2@cardiff.ac.uk

*Sarah J. Haigh, Email: sarah.haigh@manchester.ac.uk

**Author Contributions**

S.J.H., T.J.A.S. and Z. L. W supervised the project. T.J.A.S., Y.-C.W. and S.J.H. conceived the idea. Y.-C.W. conducted microscopy experiments and data analysis. G.I.L. and C.L. synthesized nanoparticles and tested electrocatalytic performance. Y.-C.W., T.J.A.S., G.I.L. and S.J.H. wrote the manuscript with inputs from all authors. All authors have given approval to the final version of the manuscript.

ACKNOWLEDGMENT

The work was supported by National Natural Science Foundation of China (Grant No.52192613), the Chinese Scholarship Council, Shuimu Scholar fellowship from Tsinghua University, the National Key R&D Project from Minister of Science and Technology (2016YFA0202704). The work was supported by EPSRC grants EP/M010619/1, EP/S021531/1 and EP/P009050/1, the European Research Council (ERC) under the European Union's Horizon 2020 research and innovation programme (Grant ERC-2016-STG-EvoluTEM-715502). The work was supported by the Henry Royce Institute for Advanced Materials, funded through EPSRC grants EP/R00661X/1, EP/S019367/1, EP/P025021/1 and EP/P025498/1 (equipment access).

# Supplementary information

## Contents

# 1. Image processing

**1.1. 2D projection image processing**

The as-collected data sets consisted of 6 stacks of HAADF STEM images with axes [x, y, temperature] and 6 stacks of EDX spectrum image data sets with axes of [x, y, energy, temperature]. The inter-frame x,y shifts between acquisitions at each temperature were calculated and minimized by cross correlating HAADF STEM image stacks according to their temperature axis. This alignment was also applied to the simultaneously acquired EDX data sets. Next, each nanoparticle was selected by manually picking the center of the particle using ImageJ[1] and cropping picked nanoparticles into a 100 x 100 pixel box using in-house python scripts.

After cropping each nanoparticle into a small square region, the cropped areas were processed using the following procedures to isolate each particle: 1) a "rolling ball" algorithm was used in ImageJ[1] with a conservative parameter (rolling ball radius 50 pixels with the sliding paraboloid option enabled) to perform a background subtraction to correct any uneven background intensity; 2) a "Huang" thresholding algorithm[2] was performed to binarize foreground particle regions; 3) a watershed segmentation algorithm was used to segment touching particles; 4) the final segmented particles were binarized again and were used as masks to extract the particle region from raw EDX data; 5) the geometric center of the segmented HAADF particle images was matched to the square region's center, the same alignments then were applied on the corresponding EDX spectrum image data; 6) using the HyperSpy python package[3], 2D EDX intensity maps were extracted for energies of 7.334–7.622 keV (Ni Kα) and 9.281–9.603 keV (Pt Lα); 7) each HAADF particle image and their EDX maps were normalized to ensure the pixel values in each image have a mean of 0 and a standard deviation of 1. The processed EDX data were used for composition analysis using standardless k-factor quantification.[4] The processed HAADF images were used for 2D area analysis and orientation matching with reference re-projections generated from reference 3D data.

**1.2. 3D reconstruction**

The 3D reference data is a HAADF STEM tomogram of about 400 PtNi nanoparticles collected using the same SPR methodology at room temperature in previous work.[5] The 3D reconstruction was then filtered by a 3D Gaussian filter with 2-pixel kernel size and aligned to its octahedral symmetry axis. A total of 400 re-projections were generated from this 3D tomogram using 2-fold rotational symmetry (C2) and including mirror portions using the EMAN2 projection generator coded[6,7] in python[8]. The orientation of the reference re-projection was described by its azimuth and altitude angles.

The matching between images of extracted nanoparticles and re-projections were performed by cross correlation. Using a 2D cross correlation implemented using the scikit-image python package[8], every processed HAADF particle image was matched with all 400 references images and their rotated forms. Re-projections were rotated from 0°–360° with 3° increments. Then, matching results were sorted according to the cross-correlation coefficient (CCC) value from high to low, with only the match giving the highest CCC value being retained. Visual inspections were conducted to discard any incorrect matches.

The circled area in Figure 4a illustrates an initial feature where HAADF intensities are lower than the surroundings, probably due to inhomogeneous etching during nanoparticle synthesis. This circled initial feature is visible in the same location during the whole heating process, showing that the nanoparticle remains at approximately the same orientation during heating and providing further confidence in the SPR assumption of a fixed orientation for individual particles throughout heating.

To perform the 3D reconstructions, the experimental images were first rotated using the in-plane rotation angle of the matched reference. Then the matched azimuth and altitude angle were used as input values for a filtered Fourier space back-projection reconstruction algorithm implemented in EMAN2,[6] with an enforced C2, C4 (Figure S8) or octahedral symmetry (Figure 3). As the orthoslices from reconstructions using different symmetries show similar morphology and intensity distribution, we conclude that the imposed symmetries do not induce artefacts and the highest symmetry (octahedral) was used for interpreting the 3D structure change of the PtNi nanoparticles for enhanced SNR. The authors also demonstrated the possibility of reconstructing asymmetric nanoparticles using the same SPR approach by automated imaging of large number particles accompanied by classification algorithms.[9] To visualize the data, the 3D tomograms were filtered with a 3D 2-pixel size Gaussian filter to reduce noise and thresholded using the 'Huang' algorithm.[2]

## 2. Consideration of microscope acquisition parameters for in-situ SPR

As the SPR methodology requires imaging many regions under the same imaging conditions and an in-situ study is usually time consuming and requires expensive instrumentation, microscope acquisition parameters should be considered based on practical limitations. There are three limitations: the specimen total beam fluence tolerance, the available acquisition time and the reconstruction quality desired.

The total tolerable beam fluence for the experiment can be calculated using the equation:

$$Total\ beam\ fluence = \frac{beam\ current \times dwell\ time}{elementary\ charge \times pixel\ size^2} \times temperature\ steps$$

The same PtNi nanoparticles were examined in previous work and were shown to damage at an electron fluence of around $5 \times 10^7$ electrons/$\text{Å}^2$.[5] Using this value as an upper limit, all electron fluence imparted to the specimen during in-situ heating experiments should be below this limit. The estimated total beam fluence used for each region during all 7 temperature steps is about $4 \times 10^5$ electrons/$\text{Å}^2$, which is much lower than the upper limit and ensures the specimen is not significantly altered by the electron beam. A pixel size of 0.386 nm was chosen to be similar to previous ex-situ spectroscopic SPR work (0.28 nm)[5] to ensure the quality of both reconstructions is comparable. The other variable parameters are pixel dwell time, temperature step and beam current. Very low electron fluence has the disadvantage of requiring a very large number of nanoparticles in the data set or producing a poor signal to noise ratio (SNR), especially for EDX spectrum images where the collection efficiency is low.

The total acquisition time required for an in-situ SPR data set can be calculated using the equation:

$$\begin{aligned} Total\ acquisition\ time \\ = dwell\ time \times frame\ size^2 \times 1.2 \times number\ of\ regions \\ \times temperature\ steps \end{aligned}$$

The factor 1.2 is an experimental factor tested on the Thermo Fisher Titan microscope and is considered as resulting from the flyback correction to move the beam from the last scanned pixel on the previous line to the first pixel of the next line. Using the equation above and current parameters, the calculated total acquisition time is estimated as 12 hours, which excludes the time for microscope alignments, manually moving between regions and switching temperatures. Increasing the dwell time or number of regions could result in obtaining better SNR images and spectrum data, or higher spatial resolution but would significantly increase the acquisition time. Reducing the number of temperature steps could reduce acquisition time but may result in the experiment failing to observe key details of the nanoparticle transformation.

Frame size and number of regions do not directly affect the total beam fluence imparted on the specimen but determine the number of particles imaged and therefore affect reconstruction quality. These two parameters are also affected by the dispersion of nanoparticles on the support. The heating chip was prepared by drop-casting nanoparticle solutions then drop-casting methanol to prevent nanoparticle agglomeration. Increasing the amount of nanoparticle solution and decreasing the amount of methanol could result in particles being packed closely, such that a smaller frame size and fewer image regions are required for the same number of particles. However, precise control of drop-casting solution is difficult. If too much nanoparticle solution is deposited onto the chip, the particles overlap and cannot be isolated for reconstruction.

3. Area identification error analysis.

Intensity line profiles at the nanoparticle surfaces were analysed to reveal the potential errors in the nanoparticle perimeter measurements. Line profiles illustrating the intensity change at the surface of three representative nanoparticles are shown in Figure S11 (the three nanoparticles are the same nanoparticles as used in Figure 1d). The HAADF intensity line profiles show high contrast differences between the nanoparticle and the background, with the identified edge marked on each image as the dashed red line. An alternative manual approach would be to mark the edge as where the intensity falls to 10% of the maximum intensity and the processed edge locations are in excellent agreement with such a manual approach (a better than 5% agreement in radial distances considering the 12 intensity profiles in Figure R2). Considering the integrated HAADF intensity, which is representative of the total number of atoms, this also shows that less than 5% of the line profile's integrated intensity is outside the areal measurement providing confidence in the accuracy of the analysis approach. An alternative perimeter measurement approach would be to assign the edge as where the HAADF intensity from the particle falls to zero in the intensity line scans. This can be expected to lead to a slight increase in the assigned area and perimeter for each particle. However, this approach has the disadvantage of being insufficiently robust in the presence of non-uniform hydrocarbon surface contamination or variable support thickness, when it creates artefacts of particles with artificially large perimeters. The analysis approach presented in Figure 2 is focused on a quantitative comparison of the particle variability and of the evolution in surface area during heating, rather than absolute perimeter quantification, thus our automated analysis was chosen to prioritize consistency and robustness.

4. Additional experimental results

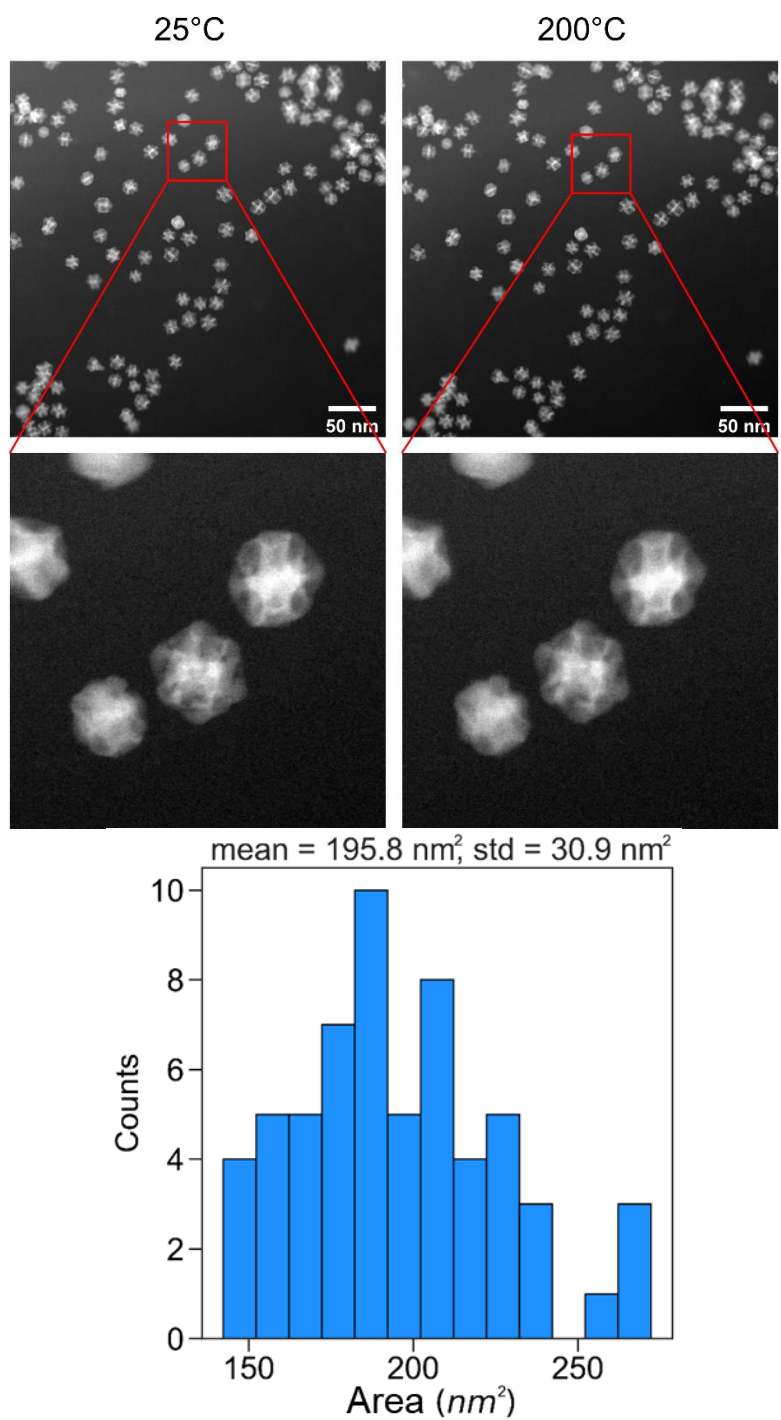

Figure S1. Nanoparticles at 25°C and 200°C show no noticeable difference in morphology. Bottom panel is the histogram of nanoparticle area distribution measured at 200°C showing a mean projected area per nanoparticle of 196 nm² and a standard deviation of 31 nm².

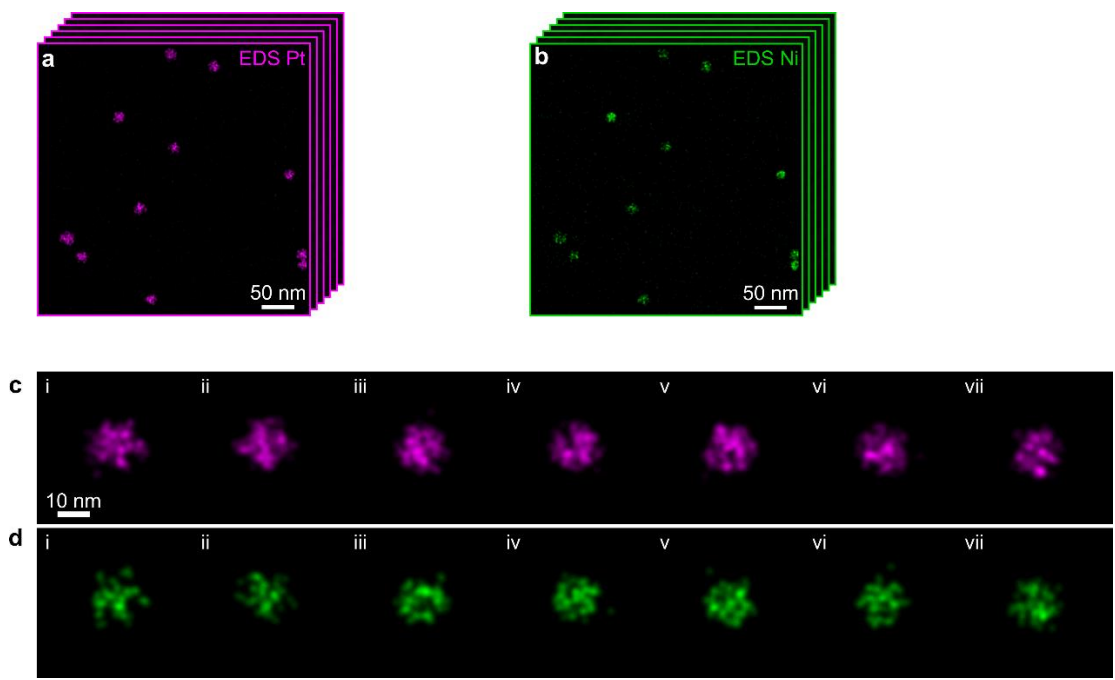

Figure S2. Overview of PtNi nanoparticles' compositional change during heating. (a) Pt and (b) Ni STEM-EDX maps simultaneously acquired with STEM-HAADF images shown in main text. (c) EDX Pt and (d) EDX Ni elemental maps showing one example nanoparticle changing from (i) 200°C to (ii) 250°C, (iii) 300°C, (iv) 350°C, (v) 400°C, (vi) 450°C, and (vii) 550°C.

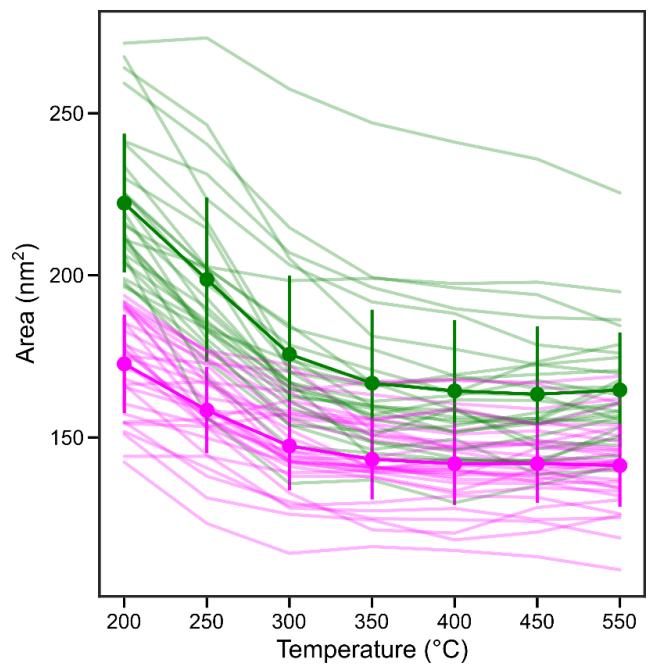

Figure S3. Changes in the 2D projected area for 60 individual nanoparticles as a function of heating temperature. Green and purple lines are used to indicate particles that are initially larger or smaller than the mean area, respectively. Bold purple (green) line is the mean for all purple (green) particles with error bars representing the standard deviation. Larger particles show a larger area reduction at the beginning of the heating (25% vs. 17% reduction for larger and smaller particles from 200°C to 350°C, respectively), likely due to the proportionally larger size of their convex surface protrusions. Nonetheless, both small and large particle data sets show similar morphological progression and areal trends: a rapid area loss from 200°C to 350°C followed by slower area loss from 350°C to 550°C. Note that the separation into large and small data sets is artificial and the nanoparticle data contains a smooth continuum with very few outliers.

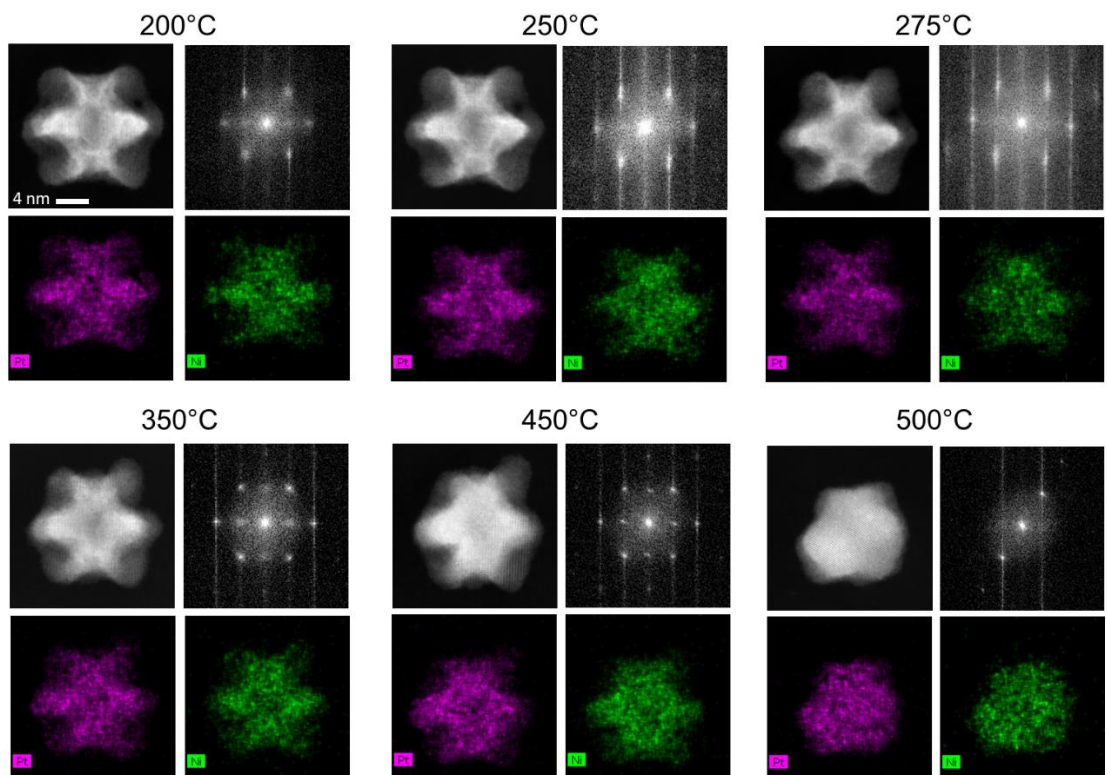

Figure S4. One example nanoparticle characterized by STEM-HAADF (top left panels) with the fast Fourier transform of the STEM-HAADF image (top right panels). The EDX Pt elemental maps (bottom left panels) and EDX Ni elemental maps (bottom right panels) are for the same area as the STEM-HAADF image. The same particle is presented after heating at 200°C, 250°C, 275°C, 350°C, 400°C, and 500°C (10 minutes at each temperature step). Scale bar at 200 °C applies to all panels.

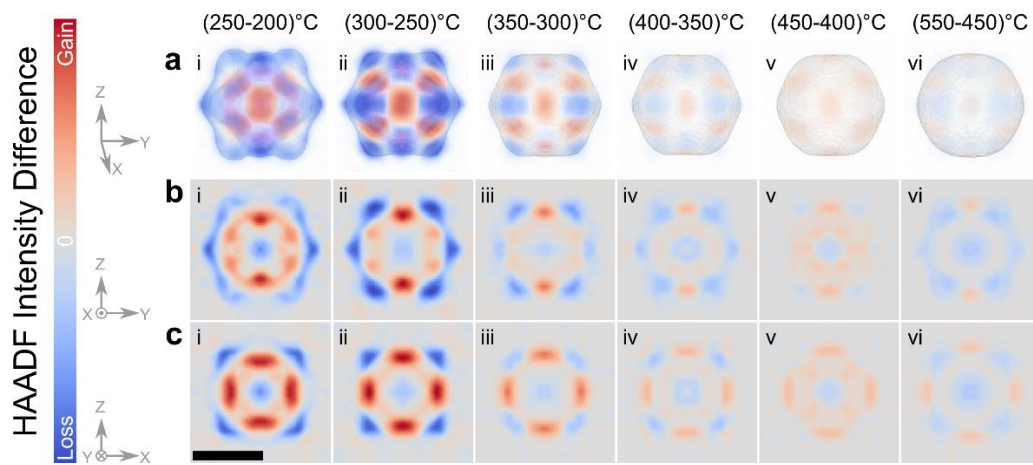

Figure S5. (a) 3D volume intensity differences between consecutive heating conditions with a transparent surface to illustrate the 3D morphology at the subsequent temperature step. (b, c) Orthoslices extracted from the center of the 3D volume intensity differences in (a) parallel to Y-Z and X-Z plane, respectively. Scale bar in (ci) is 10 nm and relates to all panels. As shown in the intensity scale, red represents an intensity gain and blue represents an intensity loss (reflecting relative enrichment or depletion of Pt respectively).

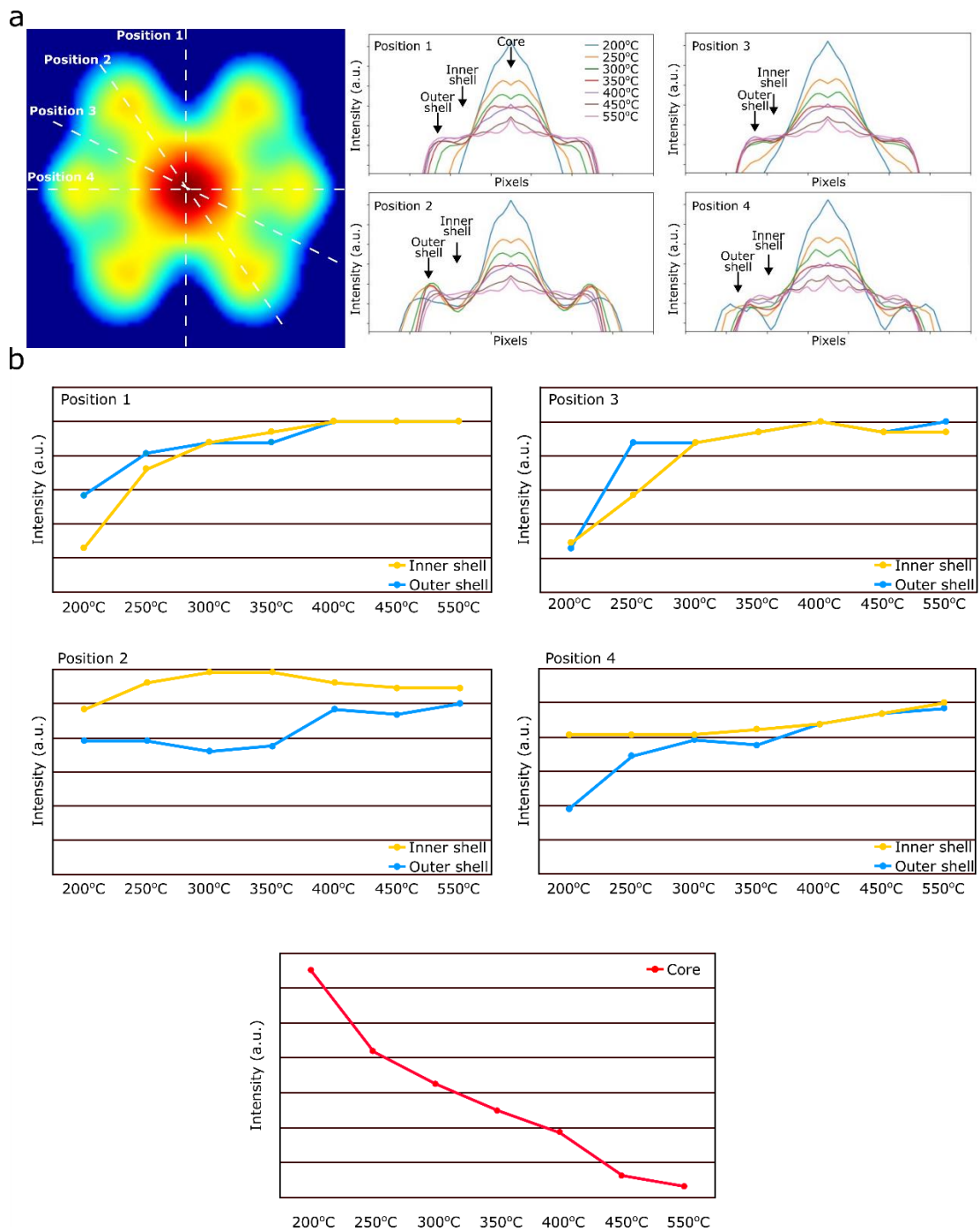

Figure S6. (a) HAADF intensity line scan from 4 positions as a function of temperature. (b) Intensities from outer shell, inner shell and core regions indicated in (a) by arrows.

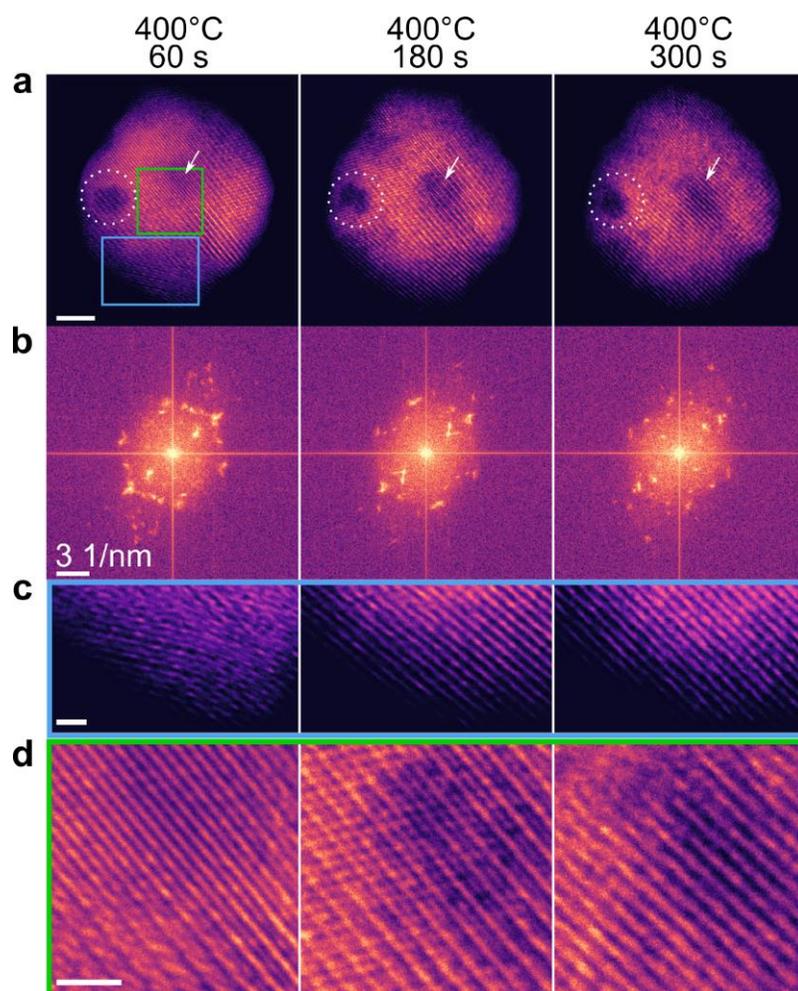

Figure S7. The morphology of the nanoparticle shown in main text Figure 4 on further heating at 400°C. The morphology does not change significantly from 60s to 300s. The circled area in (a) indicates the presence of the initial lower intensity feature seen in the same location throughout the heating process, which suggests the nanoparticle remains at the same orientation during heating. Arrows in (a) show the beam park which error resulted in a locally damaged region. Scale bars in (a), (c), and (d) are 3 nm, 1 nm, 1nm, respectively.

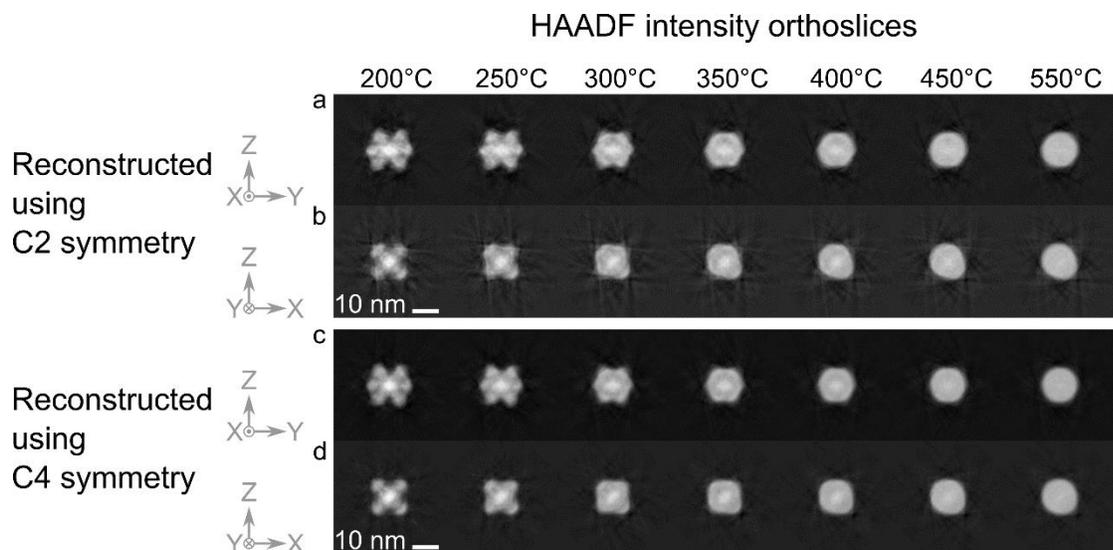

Figure S8. Orthoslices extracted from the center of 3D reconstructions using C2 or C4 symmetry respectively. The similar intensities and the morphologies of orthoslices extracted from reconstructions using different symmetries show that the imposed symmetry does not produce artefacts, but only increases SNR.

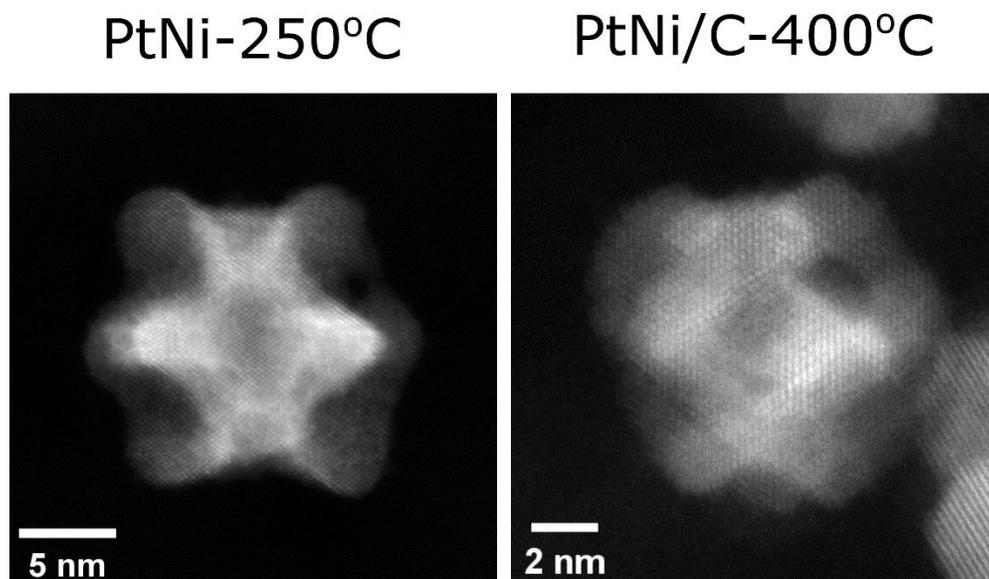

Figure S9. Comparison of STEM-HAADF images of PtNi-250$^{O}$C and PtNi/C-400$^{O}$C nanoparticles.

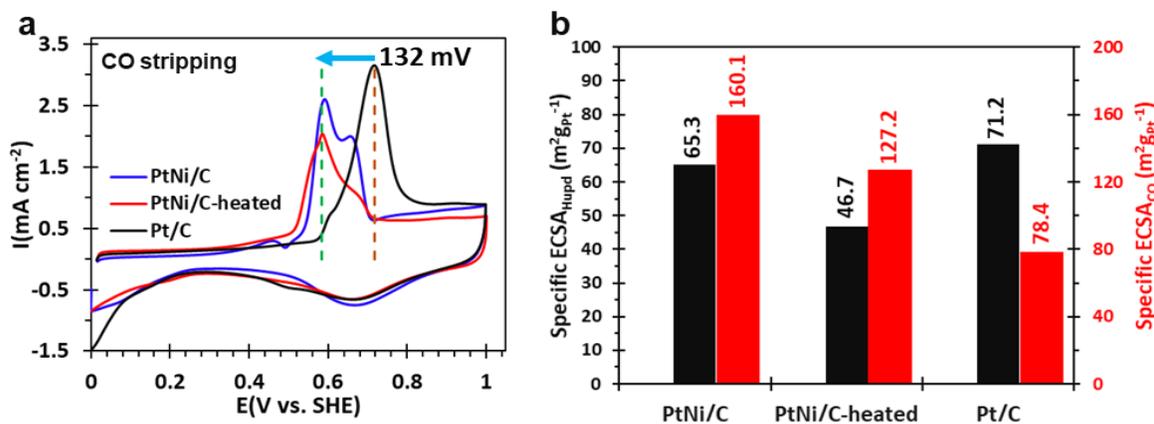

Figure S10. (a) $CO_{ads}$ electro-oxidation profiles and (b) bar graphs depicting specific $ECSA_{Hupd}$ and $ECSA_{CO}$ values.

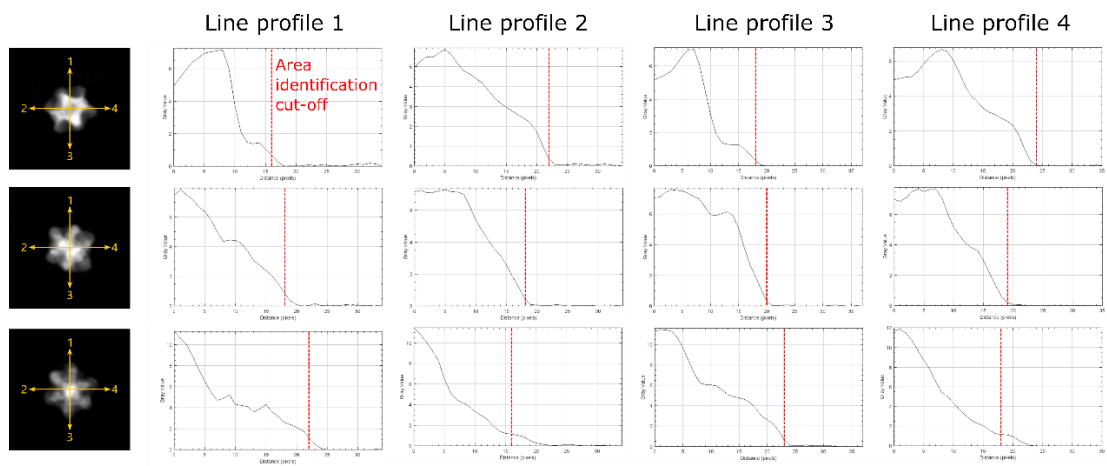

Figure S11. HAADF intensity line profiles and the identified edge identification cut-offs taken from 3 representative nanoparticles (from those in Figure 1c) to assess the accuracy of the automated particle edge identification and area calculation.

## 5. SI References